\documentclass[twocolumn,pra,superscriptaddress]{revtex4}
\usepackage{amsfonts}
\usepackage{amssymb}
\usepackage{amsmath}
\usepackage{epsfig}
\usepackage{color}
\usepackage{graphics, graphicx}
\usepackage{bbold}
\usepackage{psfrag}
\usepackage{mathcomp}
\usepackage{subfigure}
\usepackage{verbatim}
\usepackage[colorlinks,citecolor=blue]{hyperref}

\makeatletter

\newcommand{\Rmnum}[1]{\expandafter\@slowromancap\romannumeral #1@}
\makeatother

\begin{document}

\title{Effects of decoherence on diabatic errors in Majorana braiding}

\author{Zhen-Tao Zhang}
\email{zhzhentao@163.com}%
\affiliation{School of Physics Science and Information Technology, Shandong Key Laboratory of Optical Communication Science and Technology, Liaocheng University, Liaocheng 252059, People¡¯s Republic of China}
\author{Feng Mei}
\email{meifeng@sxu.edu.cn}
\affiliation{State Key Laboratory of Quantum Optics and Quantum Optics Devices,
Institute of Laser Spectroscopy, Shanxi University, Taiyuan, Shanxi 030006, China}
\affiliation{Collaborative Innovation Center of Extreme Optics,
Shanxi University, Taiyuan, Shanxi 030006, China}%
\author{Xiang-Guo Meng}
\author{Bao-Long Liang}
\author{Zhen-Shan Yang}
\affiliation{School of Physics Science and Information Technology, Shandong Key Laboratory of Optical Communication Science and Technology, Liaocheng University, Liaocheng 252059, People¡¯s Republic of China}
\date{\today}

\begin{abstract}
The braiding of two non-Abelian Majorana modes is important for realizing topological quantum computation. It can be achieved through tuning the coupling between the two Majorana modes to be exchanged and two ancillary Majorana modes. However, this coupling also makes the braiding subject to environment-induced decoherence. Here, we study the effects of decoherence on the diabatic errors in the braiding process for a set of time-dependent Hamiltonians with finite smoothness. To this end, we employ the master equation to calculate the diabatic excitation population for three kinds of decoherence processes. (1) Only pure dehasing: the scaling of the excitation population changed from $T^{-2k-2}$ to $T^{-1}$ ($k$ is the number of the Hamiltonian's time derivatives vanishing at the initial and final times) as the braiding duration $T$ exceeds a certain value. (2) Only relaxation: the scaling transforms from $T^{-2k-2}$ to $T^{-2}$ for $k=0$ and to $T^{-a}$ ($a>3$) for $k>0$. (3) Pure dephasing and relaxation: the original scaling switches to $T^{-1}$ firstly and then evolves to $T^{-2}$ in the adiabatic limit. Interestingly, the third scaling-varying style holds even when the expectation of pure dephasing rate is much smaller than that of the relaxation rate, which is attributed to the vanishing relaxation at the turning points of the braiding.

\end{abstract}

\maketitle

\section{Introduction}
Decoherence is one of the main obstacles that hinder the realization of large-scale quantum computing. The decoherence of a qubit arises from the interaction with its environment and is even worse when the number of qubits becomes larger. An alternative attractive route is to realize fault-tolerant quantum computation based on topologically protected non-Abelian anyons \cite{Kitaev01,Ivanov01,Nayak08}, where topological protection provides a natural mean to achieve fault tolerance. Non-abelian anyons can be generated in a topological superconductor system \cite{Kitaev01}, which are the zero-energy quasiparticles located at the two boundaries of the system. Such quasiparticles are now also called as Majorana zero modes (MZMs). Two MZMs and their braiding can be used to encode a topological qubit and realize topologically protected quantum gates respectively \cite{Sau10,Akhmerov10,Flensberg11,Bonderson11,Zhang13,Xue13}. Due to the non-locality nature of MZMs, the quantum information stored in the topological qubits is roubust to local perturbations \cite{Prada17,Clarke17,Deng18,Zhang18a}. The great potential of MZMs in realizing topological quantum computation have motivated numerous theoretical and experimental studies on the physical realizations of MZMs. Among them, the one-dimensional nanowire contacted with a s-wave superconductor is the most investigated system \cite{Lutchyn10,Oreg10}. Under certain condition, the nanowire system could become a topological superconductor with two MZMs in its two ends. Although some remarkable signatures of MZMs were observed in experiments recently \cite{Mourik12,Das12,Deng12,Zhang18}, the Majorana braiding still has not been demonstrated in any candidate system \cite{Alicea11,Zhu11,Aasen16}.

The braiding of a pair of MZMs is the prerequisite to verify the non-Abelian statistics as well as to realize topological quantum computation. Several methods have recently been proposed to exchange MZMs, including moving them by directly changing the positions of the boundaries of topological superconductor \cite{Alicea11}, measurement-based braiding \cite{Bonderson08}, and effectively moving MZMs via tuning the coupling between them in a network of one-dimension topological superconductor \cite{Sau11,Heck12,Hyart13,Karzig16}. The former two methods require either tuning the relevant parameters over a rather large range or reading out the states of a number of pairs of MZMs, both of which are huge challenges nowadays. In contrast, the operation needed in the third method is merely tuning the coupling strength between MZMs, which is more promising with up-to-date technique. In this work, we will discuss the practical performance of the last method in detail.

Theoretically, as the pariwise couplings between four MZMs are tuned slowly enough in a prescribed manner, the exchange of a pair of MZMs within the system could be effectively realized. However, in practice, finite operation time would bring diabatic errors to the final state of the topological qubit encoded by the two exchanged MZMs. It was proven that for a closed system the diabatic error rate is dependent on the smoothness of the time-dependent Hamiltonian at the starting and ending points. Specifically, the diabatic excitation probability scales with evolving time $T$ as $T^{-2k-2}$ \cite{Lidar09}, with $k$ as the number of the Hamiltonian's time derivatives vanishing at the initial and final times. In our case, however, the couplings between MZMs would make the system subject to its environmental bath, which causes a decoherence. Therefore, it is highly desirable to discuss how and to what extent the decoherence affects the braiding and modifies the diabatic errors.

\begin{figure}
\includegraphics[width=6cm]{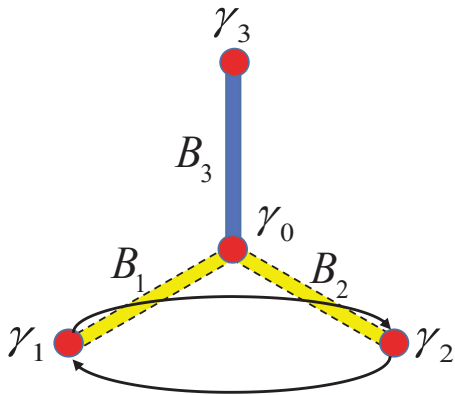}
\caption{Schematic of Y-junction. Four MZMs (denoted by solid circles) emerge at the ends of the nanowire. $\gamma_1,\gamma_2$ are the MZMs to be exchanged, and $\gamma_0,\gamma_3$ are the ancillary MZMs. $B_i$ denotes the coupling strength between $\gamma_0$ and $\gamma_i$. }
\label{Fig1}
\end{figure}

Recently, the dissipation of the MZMs coupled with an Ohmic bath have been studied \cite{Knapp16}, where it is found that the excitation population scales roughly as $T^{-2}$ in the adiabatic limit. Yet, the respective influences of the pure dephasing \cite{Knapp18} and the relaxation \cite{Song18} of the coupled MZMs on the braiding are still unclear. Most lately, Ref. \cite{Nag18} has investigated how the decoherences revise the scaling of diabatic error rate in the braiding process with infinitely smooth Hamiltonian. Interestingly, they found that the roles of the pure dephasing and the relaxation are quite different. However, their results are obtained under some special conditions, including infinitely smooth Hamiltonian and constant energy difference between the ground state and the excite state. In addition, the regime where the relaxation dominates the pure depasing has not been concerned therein. Thus, one may wonder how the decoherence affects the diabatic errors in a more general braiding process.

In this paper, we extend the study of Ref. \cite{Nag18} in three dimensions. First, a set of time-dependent Hamiltonians with finite smoothness are investigated. Based on this, we are able to compare the performances of the braiding protocols with different orders of smoothness in the presence of decoherence. Second, we consider a more practical situation in which the energy difference between the ground state and the excite state varies in the braiding process. Third, the unexplored case in the previous reference where the relaxation is stronger than the pure dephasing is included in our deliberation. Unexpectedly, we find the scaling-varying style of the diabatic errors as the pure dephasing is much weaker than the relaxation, is distinct from that of the relaxation-only case. Instead, the error scaling behaves rather similarly with that of the dephasing-dominated case. In all, the present work could improve our understanding of the impact of the decoherence on the Majorana braiding, and provide a guideline to design high-fidelity braiding protocols.

The paper is organized as follows. In Sec. 2, we introduce the model and the master equation for the Majorana system. In Sec. 3, we show the numerical results with vanishing relaxation and deduce a analytical formula to estimate the scaling of the excitation population. In Sec. 4, we numerically calculate the diabatic excitations in the presence of relaxation for both cases with and without pure dephasing, and then analyze the error scalings when the relaxation dominates the pure dephasing. We summarize the main results of this paper and discuss the limitations of the braiding time in Sec. 5.

\section{Model and master equation}

The minimal platform for braiding two MZMs is a Y-junction architecture \cite{Karzig16}, as shown in Fig. \ref{Fig1}. The MZMs $\gamma_{1,2,3}$ are localized at the far ends of the junction, while the MZM $\gamma_0$ is located at the center. The couplings between $\gamma_{1,2,3}$ and $\gamma_0$ are denoted as $B_{1,2,3}$ respectively. The braiding of the two MZMs $\gamma_1$ and $\gamma_2$ is achieved by adiabatically tuning the couplings $B_{1,2,3}$, with $B_{1,2}$ vanishing at the beginning and end. This process can be described by a time-dependent Hamiltonian
\begin{equation}\label{eq1}
  H=\sum_{j=1}^3 iB_j(t)\gamma_0\gamma_j.
\end{equation}
Without losing generality, we assume that $B_{1,2,3}$ have the same amplitude, denoted by $B_m$. In our calculations we set $B_m=1$. It is required that at least one coupling $B_j$ is zero during the braiding to guarantee the topological degeneracy associated with the MZMs. The Hilbert space of the system is four dimensional, and can be identified as a direct product of a two-fold degenerate subspace and a nondegenerate two-level subspace. The topological qubit is defined in the former subspace with the basis being the two parity states $\{|0\rangle, |1\rangle\}$. The ancillary qubit is formed in the latter subspace with the basis being the two lifted parity states: the ground state and the excite state $\{|g\rangle,|e\rangle\}$.

To implement the braiding, the Hamiltonian should be varied adiabatically in three steps, as plotted in Fig. 2. If the process is completely adiabatic, the initial state $|g\rangle(|0\rangle+|1\rangle)$ will evolve to $|g\rangle(|0\rangle+i|1\rangle)$ after the braiding. Note that the overall parity $P=\gamma_0\gamma_1\gamma_2\gamma_3$ is conserved in the braiding process due to its commutation relation with the Hamiltonian. This means that an diabatic excitation of the ancillary qubit is accompanied by a bit-flip error of the topological qubit. Hence, the possibility of the diabatic errors is directly related to the fidelity of the braiding operation. Due to the similarity of the three steps of the braiding process, we only investigate the first step with duration $T$. Projecting the Hamiltonian into the even parity subspace via the transformations $\sigma_{x,y,z}=i\gamma_0\gamma_{1,2,3}$, the Hamiltonian could be expressed as:
\begin{equation}\label{eq2}
  H_e=\vec{B}\cdot\vec{\sigma}
\end{equation}
where $\vec{B}=(B_1,B_2,B_3)$, $\vec{\sigma}=(\sigma_x,\sigma_y,\sigma_z)$ with $\sigma_{x,y,z}$ being the Pauli operators.

\begin{figure}
\includegraphics[scale=0.37]{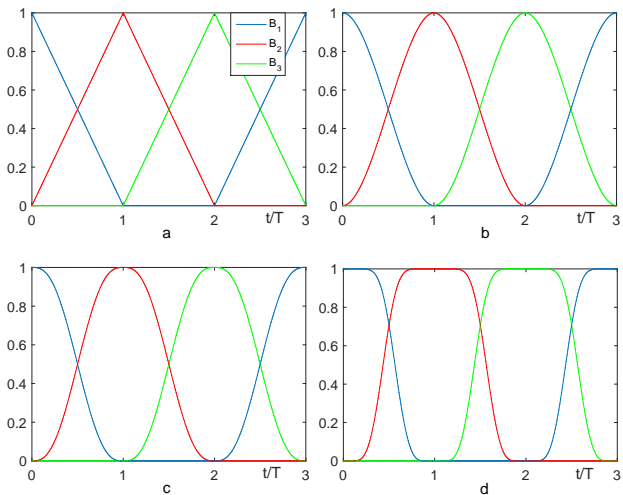}
\caption{Time dependent of MZMs couplings for braiding $\gamma_1$ and $\gamma_2$ as shown in Fig. 1. a-c are k-order derivative continuous $\vec{B}$ with k=0,1,2 respectively. For comparison, the infinitely smooth $\vec{B}$ is illustrated in d. The braiding process consists of three steps. In each step there exists one component of $\vec{B}$ vanishing throughout to maintain the topological degeneracy. The norm of $\vec{B}$ varies with time in a-c while the quantity is constant in d.}
\end{figure}

Derivative discontinuities in the time-dependence of $\vec{B}$ are known to induce diabatic errors in the topological qubit, whose possibility scales as the power-law $T^{-2k-2}$ for the isolated system. Thus, the error rate will descend more quickly with $T$ for smoother $\vec{B}(t)$ at the initial and final times. For simplicity, we rescale the time $t$ to the unitless quantity $s=t/T$. In the later calculations, we adopt the interpolating function of $\vec{B}$ given by the regularized incomplete beta function $\theta_k(s)$ \cite{Rezakhani10,Albash15}
\begin{equation}
  \theta_k(s)=\frac{D_s(1+k,1+k)}{D_1(1+k,1+k)}
\end{equation}
in which $D_s(a,b)=\int_0^sy^{a-1}(1-y)^{b-1}dy$, with $Re(a)$, $Re(b)>0$, and $|s|\leq1$. In the first step, the vector $\vec{B}=(1-\theta_k(s),\theta_k(s),0)$. In Fig. 2a-c, the components $B_1, B_2, B_3$ in the whole braiding process are illustrated for k=0, 1, 2, respectively. For comparison, we have drawn the vector $\vec{B}$ which is infinitely smooth at the turning points of each step, as shown in Fig. 2d.\\
\indent To take account of the interactions with a parity-conserving bath, we introduce a master equation for the braiding process. Firstly, we assume that the operators of the system $\sigma_{x,y,z}$ are coupled to the bath with the coupling strengthes proportional to $B_{1,2,3}$, respectively. This is consistent with the fact that when $\gamma_0$ and $\gamma_i$ are uncoupled, the quantity $\sigma_i=i\gamma_0\gamma_i$ is non-local and will not coupled to the local bath. Thus, the interaction Hamiltonian reads
\begin{equation}\label{eq3}
  H_I=\sum_is_iB_i\sigma_i\cdot\hat{Q}_i
\end{equation}
in which $s_iB_i$ is the coupling strength between $\sigma_i$ and the bath operator $\hat{Q}_i$. Under Born approximation and Markov approximation, the master equation of the open MZMs system could be addressed in the Lindblad form as \cite{Nag18}
\begin{eqnarray}
  \epsilon\dot{\rho}(s)=-i[H_e,\rho(s)]+\alpha(s)[\tau_z\rho(s)\tau_z-\rho(s)]\nonumber\\+\beta(s)[\tau_-\rho(s)\tau_+-\frac{1}{2}(\tau_+\tau_-\rho(s)+\rho(s)\tau_+\tau_-)],
\end{eqnarray}
where $\epsilon=1/T$, $\alpha$ and $\beta$ are the pure dephasing rate and the relaxation rate respectively. $\tau_z$ is the Pauli operator which is diagonalized in the instantaneous energy eigenstates basis, and $\tau_+(\tau_-)$ is the raising(lowering) operator. Here we have assumed that the bath is in the low-temperature limit, and direct excitations of the system by the bath are exponentially suppressed. Because of the $\vec{B}$-dependence of the interaction Hamiltonian, the values of $\alpha$ and $\beta$ rely on $\vec{B}$, and in turn are functions of time $s$. Through some algebraic deduction, we get the relations:
\begin{eqnarray}
  \alpha(s)&=&\eta_0[s^2_xB^4_x(s)+s^2_yB^4_y(s)]/|\vec{B}|^2,\label{a}\\
  \beta(s)&=&\frac{\eta}{4}(s^2_x+s^2_y) B^2_x(s)B^2_y(s)/|\vec{B}|^2 \label{b}
\end{eqnarray}
 where the factors $\eta_0, \eta$ are determined by the microscopic properties of the bath and assumed to be invariant during the braiding process. The norm $|\vec{B}|$ represents the energy difference between the ground state and the excite state, which is not necessarily constant. For the sake of analysis, we transform the master equation to the Bloch equation in terms of a Bloch vector $\vec{R}=(r_x,r_y,r_z)$,
\begin{equation}\label{eq4}
  \epsilon\dot{\vec{R}}=2[\vec{B}\times\vec{R}+(\alpha-\beta)\vec{B}\times(\vec{B}\times\vec{R})/|\vec{B}|^2-2\beta(\vec{B}/|\vec{B}|+\vec{R})]
\end{equation}
where the time derivatives are referred to $s$.
\section{Effect of pure dephasing}
Now we investigate how the excitation populations are influenced by the pure dephasing. For this purpose, we set the relaxation rate $\beta$ to be $0$ in this section. Since $T$ is always finite, the states of system during the braiding should be superpositions of the ground state and the excite state. Therefore, the pure dephasing of the system would affect the scaling of the excitation population with $T$. To clarify this issue, we will firstly illustrate the numerical results of the Bloch equation, and then give an analytical interpretation.\\
\begin{figure}
\includegraphics[width=8.5cm,height=5.5cm]{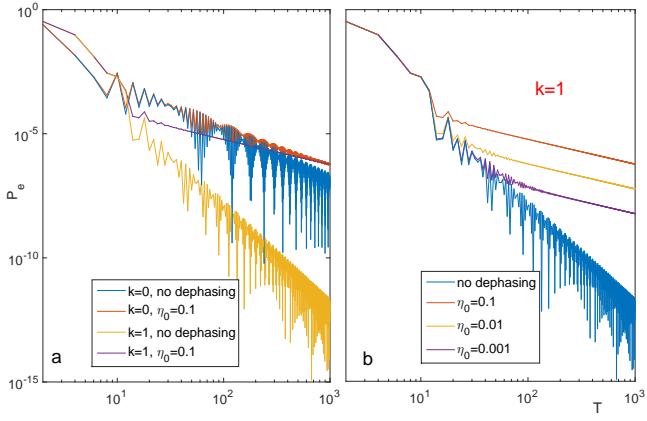}
\caption{Numerical results of the excitation populations $P_e$ with and without pure dephasing. a. $P_e$ for k=0 with (blue line) and without (red line) the dephasing, and k=1 with (yellow line) and without (purple line) the dephasing; b. $P_e$ for k=1, with the dephasing factor $\eta_0=0$ (blue line), 0.001(purple line), 0.01(yellow line), 0.1(red line). The other parameters are set as: $s_x=s_y=0.1$, $\alpha=0$. }
\end{figure}
\subsection{Numerical results}
We numerically solve the Bloch equation and show the excitation populations $P_e$ varying with $T$ in Fig. 3. Specifically, in Fig. 3a, $P_e$ was calculated for k=0, 1 with and without the pure dephasing. In the absence of the dephasing, the excitation population decreases with $T$ and obeys the power-law $T^{-2k-2}$ as expected. Therefore, if $\vec{B}$ varies more smoothly at the turning points of each step, the descending of the excitation population with the step duration will be faster. However, this picture would be interrupted by the pure dephasing. From Fig. 3a, we can see that the scaling of $P_e$ is necessarily turned from $T^{-2k-2}$ to $T^{-1}$ for $k=0,1$ (larger $k$ not shown here) at certain $T$. Moreover, the value of $T$ where the turning occurs is determined by the dephasing factor $\eta_0$ as the parameters $s_x,s_y$ are fixed (see Fig. 3b). Qualitatively, stronger dephasing will lead to the turning happened at a shorter $T$. Notice that this feature is also found in the braiding process with infinitely smooth $\vec{B}$ \cite{Nag18}. Based on these results, we can state that no matter how smoothly the Hamiltonian is tuned, the power-law $T^{-2k-2}$ of the diabatic error will uniquely transform to $T^{-1}$ as $T$ is large enough. In other words, the power-law $T^{-2k-2}$ only play its role in a limited range of $T$, whose upper boundary is decided by the strength of the pure dephasing. It is worth noting that the dephasing rate $\alpha$ varies with time in the braiding process, therefore it is impossible to define an overall pure dephasing time.
\subsection{Analytical derivation}
To reveal the derivation of the dephasing induced $T^{-1}$ scaling of the excitation population, we rewrite the Bloch equation in a more compact manner after seting $\beta=0$, i.e.,
\begin{equation}\label{eq5}
  \dot{\vec{R}}=M\vec{R}
\end{equation}
where $M=2(A+S)$, $S=\alpha A^2/|\vec{B}|^2$ and $A$ is the matrix
\begin{equation}
  A=\left(
      \begin{array}{ccc}
        0 & 0 & B_y \\
        0 & 0 & -B_x \\
        -B_y & B_x & 0 \\
      \end{array}
    \right).
\end{equation}
The solution of the Bloch equation in the above form admits an adiabatic series expansion written as:
\begin{equation}\label{eq6}
 \vec{R}=\vec{R}_0+\epsilon\vec{R}_1+\epsilon^2\vec{R}_2+\cdots
\end{equation}
 where $\vec{R}_0=-\vec{B}/|\vec{B}|$ denotes the Bloch vector of the instantaneous ground state. Substituting the above formula into Eq. \ref{eq5} and equating both sides of the equation at each order in $\epsilon$, we can obtain the $j$-th order correction for $\vec{R}$
\begin{equation}\label{eq7}
 \vec{R}_j=f_{j-1}\vec{R}_0+M^{-1}\dot{\vec{R}}_{j-1}
\end{equation}
with
\begin{equation}
f_{j-1}(s)=\int_0^1 ds \dot{R}_0^TM^{-1}\dot{R}_{j-1}.
\end{equation}
 According to the definition, $\vec{R}_0$ is orthogonal to $M^{-1}\dot{\vec{R}}$. Thus, we can split $\vec{R}$ into two components
\begin{equation}\label{eq8}
 \vec{R}=f(s)\vec{R}_0+\vec{R}_\bot
\end{equation}
where $\vec{R}_0\bot\vec{R}_\bot$,
\begin{eqnarray}
f(s)=1+\epsilon f_0(s)+\epsilon^2f_1(s)+\cdots\nonumber\\
\vec{R}_\bot=M^{-1}(\epsilon\dot{\vec{R}}_0+\epsilon^2\dot{\vec{R}}_1+\cdots).
\end{eqnarray}
The excitation population is related to the Bloch vector difference $\vec{R}(1)-\vec{R}_0(1)$, so we first derive it before calculating $P_e$. Making use of Eq. \ref{eq8}, we get
 \begin{eqnarray}
\vec{\delta}=\vec{R}(1)-\vec{R}_0(1)&=&f(1)\vec{R}_0+\vec{R}_\bot-\vec{R}_0(1)\nonumber\\
                         &=&(f(1)-1)\vec{R}_0(1)+\vec{R}_\bot(1)
\end{eqnarray}
\indent We concern two limits: infinite smooth Hamiltonian and closed MZMs system. In the former case, $P_e\propto (f(T)-1)$, in the latter $P_e\propto|\vec{R}_\bot|^2$. For the general situation where dephaing is present and the Hamiltonian is finite smooth, the diabatic excitation population can be expressed as
\begin{eqnarray}\label{eq9}
 P_e&=&c_1(f(1)-1)+c_2|\vec{R}_\bot(1)|^2\nonumber\\&=&c_1(\epsilon f_0(s)+\epsilon^2f_1(s)+\cdots)+c_2|\vec{R}_\bot(1)|^2
\end{eqnarray}
where $c_{1,2}$ are $T$-independent constants. The first term can be approximated to the lowest order of $\epsilon$ with the prefactor $c_1f_0(1)$ depending on the dephasing rate $\alpha$. The second term relies on the smoothness of the Hamiltonian and is proportional to $\epsilon^{2k+2}$. When $T$ is so small that the second term overwhelms the first one, $P_e$ decreases as quickly as $\epsilon^{2k+2}$. With the increasing of $T$ the effect of the dephasing appears, and the scaling will gradually convert to $\epsilon$ for any $k$. Moreover, since the first term of Eq. \ref{eq9} has little relation with $k$, the diabatic excitations after the conversion can not be sizably reduced by solely improving the smoothness of the Hamiltonian. On the other side, a suppressed dephasing would lead to an expanded $T$ range of the $\epsilon^{2k+2}$ scaling, which is profitable for implementing the braiding with high fidelity in a shorter duration, just as shown in Fig. 3b. In a word, the appearance of the error scaling $\epsilon$ for any smoothness of the Hamiltonian is the consequence of the interplay of the diabatic excitation and the pure dephasing.
\section{Effects of decoherence including relaxation}
In this section, we will further study the diabatic error in the presence of relaxation. To this end, we calculate the scaling of the excitation population without and with the pure dephasing. The result of the former case is shown in Fig. 4a. It can be found that the excitation population does not have a universal scaling for different $k$. For $k=0, 1$ there is no obvious scaling transformation as $T$ increases. Basically, the scaling is always $\epsilon^2$ for $k=0$ and $\epsilon^4$ for $k=1$. However, if $k>1$, the scaling is subjected to an remarkable conversion from $\epsilon^{2k+2}$ to $\epsilon^a$ with the exponent $3\leq a<4$. After the conversion, the exponent would ascend slowly with $T$. Obviously, the diabatic excitation in this case decreases more quickly than that in the dephasing-only case after their scaling conversions. We can conclude that the pure dephasing modifies the excitation population in a severer manner compared with the relaxation.\\
\begin{figure}
\includegraphics[scale=0.32]{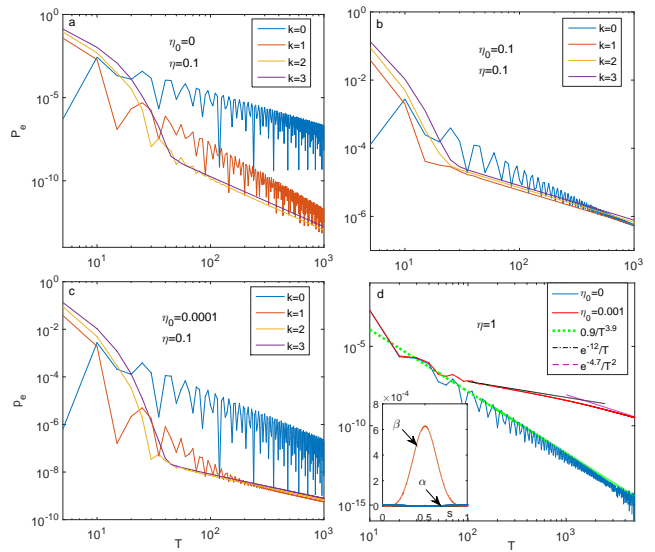}
\caption{Numerical results of the excitation populations with finite relaxation rate. a-c the excitation population for k=0-3 with the relaxation rate $\eta=0.1$ and the dephasing rate $\eta_0=0, 0.1, 0.0001$ respectively. d. the excitation population for $k=1$ with $\eta=1$ and $\eta_0=0$ (blue solid line), $\eta_0=0.001$ (red solid line). The former is fitted with the function $0.9T^{-3.9}$ (dotted green line), and the latter is fitted with the function $e^{-12}T^{-1}$ (dot-dashed black line) in the middle range of $T$, $e^{-4.7}T^{-2}$ (dashed purple line) in the final range. Inset: the evolution of the dephasing rate $\alpha$ and the relaxation rate $\beta$ during the first step of the braiding. The other parameters are same with those in Fig. 3.}
\end{figure}
Now we turn to the generic situation: both the relaxation and pure dephasing are nonvanishing. It was proved in Ref. \cite{Nag18} that for the infinite smooth Hamiltonian, when the dephasing rate $\alpha$ is constantly larger than the relaxation rate $\beta$, the diabatic error decreases as quickly as $e^{-\sqrt{T}}$ in relative small-$T$ range, and then $1/T$ in the middle range, and finally $1/T^2$ in the adiabatic limit. In the context of finite smooth Hamiltonian, the scaling varies with $T$ in a similar style, as shown in Fig. 4b-d. In the small-$T$ range, the decoherence has little effect on the evolution of the MZMs system, so $P_e$ is reduced following the scaling $\epsilon^{2k+2}$. As the duration is large enough so that the dephasing plays a part in the evolution, the scaling of the diabatic error would transform to $\epsilon$. When $T$ is further expanded so that the relaxation starts to work, the scaling is approaching $\epsilon^2$. Surprisingly, we find the basic pattern of the varying scaling can be extended to the regime $\overline{\alpha}\ll\overline{\beta}$ ($\overline{\alpha}$, $\overline{\beta}$ denote the expectations of $\alpha$ and $\beta$ respectively), see Fig. 4c-d. In Fig. 4d, we fit the curves of $P_e$ with $k=1$ and compare the scaling for weak dephasing $\eta_0=0.001$ and vanishing dephasing $\eta_0=0$. As $T$ is small, there are no difference between the two cases. With the enlarging of $T$, the curves start to bifurcate: the scaling of the weak dephasing curve turns to $\epsilon$ and varies gradually up to $\epsilon^2$, while the second curve roughly preserves the original scaling, fitted to be $\epsilon^{3.9}$, see Fig. 4d. This result is seemingly counter-intuitive, since one may naively expect that the scaling in the strong relaxation and weak dephasing regime is approaching to that in the zero dephasing and finite relaxation regime. \\
\indent In order to catch the underlying physics of the weird phenomenon, we compare $\beta$ with $\alpha$ in the whole process. According to Eq. \ref{a} and \ref{b}, we have calculated the two rates as function of $s$, as shown in the inset of Fig. 4d. Specially, we find $\beta=0$, and $\alpha\neq0$ at the beginning and ending of the step. This happens because at the turning points of the braiding only two MZMs are coupled together and interact with the environment, which results in the vanishing relaxation \cite{Song18}. Therefore, while the relaxation rate far exceed the pure dephasing rate in most of the time, the latter overtake the former near the turning points. It is reasonable to think that the outperform of the dephasing relative to the relaxation near the turning points makes the scaling follow the style of $\epsilon^{2k+2}-\epsilon-\epsilon^2$, just like the case with $\alpha>\beta$ for all $s$. Thus, although relaxation could make the diabatic error decrease as quickly as $\epsilon^a$ with $a>3$ for $k>0$, a bit of pure depasing could slow down the decrease by modifying the scaling to $\epsilon-\epsilon^2$. \\
\section{Summary and Discussions }
\indent In summary, we have studied the effects of decoherence on the MZMs braiding with general time-dependent Hamiltonians for three kinds of decoherence processes. When the decoherence is pure dehasing, the scaling of the excitation population changes from $T^{-2k-2}$ to $T^{-1}$ as the braiding duration $T$ exceeds a certain value, no mather how large $k$ is. When the decoherence is merely relaxation, there does not exist an uniform scaling. For $k=0,1$, the scaling remains as $T^{-2k-2}$, and no sharp scaling conversion occurs. In contrast, for $k>1$, a conversion from $T^{-2k-2}$ to $T^{-a}$ ($a>3$) takes place as $T$ reaches the threshold value. Lastly, as the pure dephasing coexists with the relaxation, the original scaling switches to $T^{-1}$ at first, and then evolves to $T^{-2}$ in the adiabatic limit. Besides, we find the third scaling-varying style holds even when the expectation of pure dephasing rate is much smaller than that of the relaxation rate, which can be attributed to the vanishing relaxation at the turning points of the braiding. Our study resolves and distinguishes the different roles of the pure dephasing and relaxation of the system in the braiding process with finite-smoothness Hamiltonians. Roughly speaking, the pure dephasing is more detrimental to the braiding operation than the relaxation.

From our results, we know that the decoherence could prolong the progress to the ideal adiabatic braiding. Under the practical situation, establishing the optimal braiding time to achieve the highest fidelity is a key issue in topological quantum computation. Now we discuss some limitations of the operation time. In our calculations, the unit of $T$ is the inverse of $B_m$, the maximum of the MZMs coupling strength. Thus, the operation time needed to gain a required fidelity is inversely proportional to $B_m$, whose upper limit is the superconducting gap $\Delta$ of the host of the MZMs. Actually, $B_m$ should be small enough compared with the gap. The reason is that if the ingap bound states are two close to the gap edge, they would be excited to the continuum states distributed outside the gap via Landau-Zener transitions. Therefore, the operation time and the value of $B_m$ should be balanced to minimize the errors induced by the diabatic excitations and the continuum excitations. At this moment, the effects of decoherence to the diabatic errors should be concerned. In a word, these factors determine the lower limit of the operation time. On the other hand, it is not allowed to extend the braiding process unlimitedly. This is because the topological qubit also suffers from parity-breaking environments \cite{Cheng12,Budich12,Rainis12,Schmidt12}, such as quasiparticle poisoning \cite{Karzig17}. They would directly change the parity of the topological qubit and lead to a finite lifetime of the qubit. Thus, the whole braiding duration should be shorter than the lifetime. The optimal braiding time could be obtained after taking all the above error sources into account.

\section{Acknowledgments}
We thank the very helpful discussions with Dong E. Liu. ZTZ is funded by the National Nature Science Foundation of China (No.11404156) and the Startup Foundation
of Liaocheng University (No.318051325). FM is founded by the National Key R\&D Program of China
(2017YFA0304203); National Nature Science Foundation of China (No.11604392, 11434007); The Startup Foundation of Shanxi University; Changjiang Scholars and Innovative Research Team in University of Ministry of Education of China (PCSIRT)(IRT\_17R70); Fund for Shanxi 1331 Project Key Subjects Construction; 111 Project (D18001). ZSY is founded by Natural Science Foundation of Shandong Province (ZR2018MA044).

\end{document}